\documentstyle[12pt]{article}
\input psfig
\def\reference{\parskip 0pt\par\noindent\hangindent 0.5 truecm}

\def\deg{$^{\circ}$}
\def\apj{ApJ\ }

\def\baas{BAAS\ }
\def\mnras{MNRAS\ }

\begin{document}
%
%
\title{The shroud around the ``compact, symmetric''\break
radio jets in NGC~1052}
%


\author{R.C. Vermeulen$^{1}$,
 E. Ros$^{2}$,
 K.I. Kellermann$^{3}$, \\
 M.H. Cohen$^{4}$,
 J.A. Zensus$^{2,3}$ and
 H.J. van Langevelde$^{5}$
} 

\date{}
\maketitle

{\center

$^1$       Netherlands Foundation for Research in Astronomy, P.O. Box
           2, NL--7990 AA Dwingeloo, The Netherlands. 
           rvermeulen@astron.nl\\[3mm]
$^2$       Max-Planck-Institut f\"ur Radioastronomie,
           Auf dem H\"ugel 69. D-53121 Bonn, Germany. 
           ros@mpifr-bonn.mpg.de; azensus@mpifr-bonn.mpg.de\\[3mm]
$^3$       National Radio Astronomy Observatory, 520 Edgemont Road,
           Charlottesville,\break VA 22903, U.S.A. 
           kkellerm@nrao.edu\\[3mm]
$^4$       California Institute of Technology, Pasadena, CA 91125, U.S.A. 
           mhc@astro.caltech.edu\\[3mm]
$^5$       Joint Institute for VLBI in Europe, P.O. Box 2, NL--7990
           AA Dwingeloo,\break The Netherlands. 
           langevelde@jive.nl\\[3mm]
}

%
\begin{abstract}

  This is a paper on young jet material in a frustratingly complex
  environment.

  NGC\,1052 has a compact, flat- or GHz-peaked-spectrum radio nucleus
  consisting of bi-sym\-metric jets, oriented close to the plane of the
  sky. Many features on both sides move away at $v_{\rm app}\sim0.26c$
  (${\rm H}_0=65$\,km\,s$^{-1}$\,Mpc$^{-1}$). VLBI at seven frequencies
  shows a wide range of spectral shapes and brightness temperatures;
  there is clearly free-free absorption, probably together with
  synchrotron self-absorption, on both sides of the core. The absorbing
  structure is likely to be geometrically thick and oriented roughly
  orthogonal to the jets, but it is patchy.

  H{\sc i} VLBI shows atomic gas in front of the approaching as well as
  the receding jet. There appear to be three velocity systems, at least
  two of which are local to the AGN environment. The ``high velocity
  system", 125 to 200\,km\,s$^{-1}$ redward of systemic, seems
  restricted to a shell 1 to 2\,pc away from the core. Closer to the
  centre, this gas might be largely ionised; it could cause the
  free-free absorption.

  WSRT spectroscopy shows 1667 and 1665 MHz OH absorption over a wide
  velocity range. OH and H{\sc i} profile similarity suggests
  co-location of molecular and atomic ``high velocity'' gas; the
  connection to H$_2$O masing gas is unclear. Further, at ``high
  velocity'' we detected the OH 1612\,MHz satellite line in absorption and
  the 1720\,MHz line in emission, with complementary strengths.

\end{abstract}

{\bf Keywords:}
galaxies:active,\ galaxies:NGC\,1052,\ galaxies:jets,\ galaxies:nuclei

\bigskip

%
%
%
%
%
%
%
%
%
%
%
%

\section{Introduction}\label{intro}

The nearby\footnote{We adopt ${\rm H}_0$=65\,km\,s$^{-1}$\,Mpc$^{-1}$, with
no corrections for local deviations from the Hubble flow, so that the
optical stellar absorption line heliocentric redshift,
$cz=1474$\,km\,s$^{-1}$ (Sargent et al.\ 1977), corresponds to a
distance of 22\,Mpc and implies that 1\,mas $\approx$ 0.1\,pc} LINER
(e.g., Gabel et al.\ 2000) elliptical galaxy NGC\,1052 has an unusually
prominent central radio source (1 to 2\,Jy). It is variable on timescales
of months to years (e.g.\ Heeschen \&\ Puschell 1983) and has a fairly
flat spectrum between 1 and 30\,GHz, which has sometimes been classified
as Gigahertz peaked (e.g., de Vries et al.\ 1997). The overall radio
structure is core-dominated, and has two lobes spanning only
about 3\,kpc (Wrobel 1984), so that NGC\,1052 meets the traditional
size limit for CSSs, but not for CSOs.  With VLBI, detailed sub-parsec
scale scrutiny of the active nucleus and its inner environment are
possible.

A more extensive analysis of the data presented below is in Vermeulen
et al.\ (2002).

\section{Kinematics}\label{kinem}

\begin{figure}[ht!]
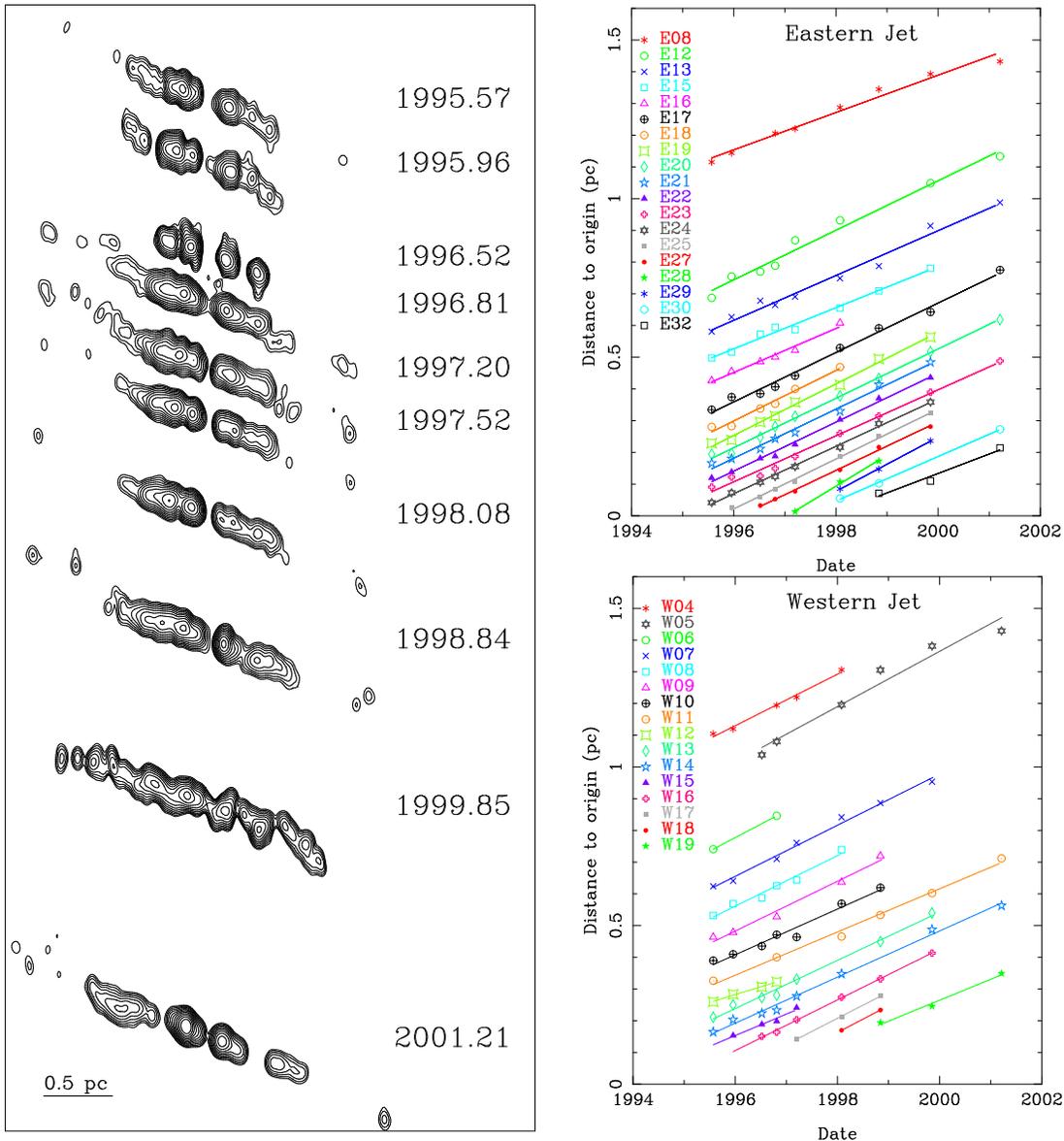

    \begin{center}
    \psfig{file=vermeulen_fig01a.ps,height=15.5cm}
    \null\vskip -16.5 truecm\null\hskip 7 truecm
    \psfig{file=vermeulen_fig01b.ps,height=7.75cm}
    \null\vskip -0.5 truecm\null\hskip 6.9 truecm
    \psfig{file=vermeulen_fig01c.ps,height=7.75cm}
    \caption{
      Left: VLBA 15\,GHz contour images of NGC\,1052 at ten
         epochs, aligned following the kinematic analysis of
         Sect.\,\ref{kinem}, and
         shown spaced by their relative time intervals. All contour levels
         increase by factors of $3^{1/2}$, starting at 0.58\%\ of the
         peak brightness in each image. The restoring beams are
         all 0.1$\times$0.05\,pc in P.A.\ 0\deg.\hfill\break
      Right: Distances to the adopted origin for the Gaussian
         components (see Sect.\,\ref{kinem}) fitted to these datasets.
         Lines show the best-fit linear velocities.
            }
\label{tenims}
\end{center}
\end{figure}

Ten epochs of 15\,GHz VLBA data (Fig.\,\ref{tenims}) show a two-sided
source, with oppositely directed, slightly curved jets and a prominent
gap 0.1 to 0.2\,pc west of the brightest feature in most images. At our
high linear resolution, NGC\,1052 shows complex evolution, which we
have quantified using 
a consistent set of
moving components,
rank-numbered ``E''ast and ``W''est by decreasing distance to the centre
(Fig.\,\ref{tenims}). While not every one of these should be
interpreted as a true physical entity (a plasmon or a ``cannon-ball''),
we are confident that the set as a whole gives an appropriate
representation of the motions in the jets. There is a plausible
relative alignment of the epochs, in which the motions of most of the
components, on both sides of the centre, are consistent with being
ballistic: linear on the sky and constant over time. Features on the
two sides move in opposite directions with roughly equal 
apparent
velocities of $0.26\pm0.04\,c$. Using these error margins, the
jets are oriented at most 33\deg\ from the plane of the sky.

\section{Ionised gas}\label{ionis}

\begin{figure}[ht!]
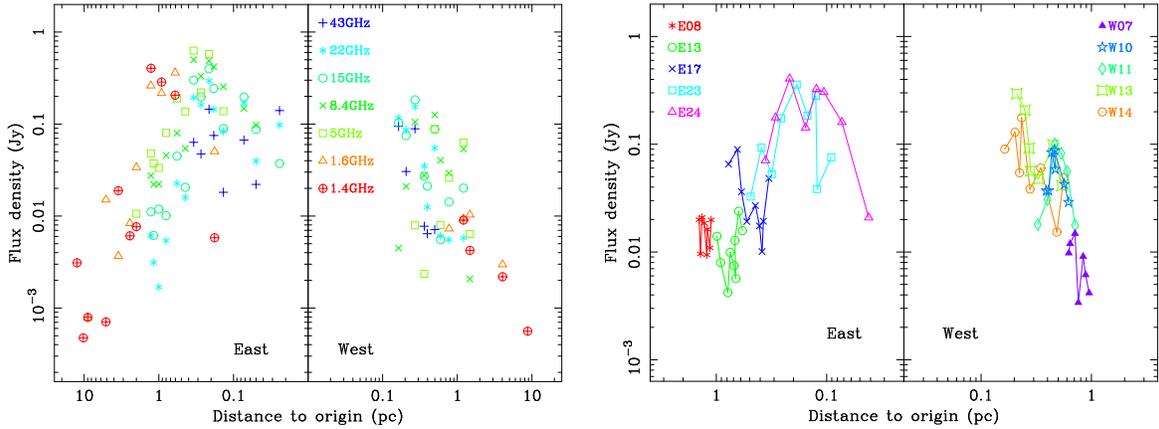

    \begin{center}
    \null\vskip 0.001 truecm\null\hskip -7.9 truecm
    \psfig{file=vermeulen_fig02a.ps,height=5.6cm,angle=-90}
    \null\vskip -6.15 truecm\null\hskip 7.8 truecm
    \psfig{file=vermeulen_fig02b.ps,height=5.6cm,angle=-90}
    \caption{
         Left: All flux densities from a multi-frequency
            multi-component VLBI analysis. The
            diversity of spectral shapes is
            emphasised by the frequency colour coding.\hfill\break
         Right: Flux density changes with time for some Gaussian
            components fitted to the multi-epoch 15\,GHz observations
            shown as a function of their distance to the origin.
            }
\label{ffflux}
\end{center}
\end{figure}

The left panel in Fig.~\ref{ffflux} shows data at 43, 22, 15, 8, 5,
1.6, and 1.4\,GHz from VLBA observations within a few days of each
other in July 1997. There is a wide range of spectral shapes, which
proceed from steep, through convex, to highly inverted, from the outer
jets towards the middle, and produce a distinctive central hole in the
VLBA images. On both sides of the gap there are components with a
low-frequency brightness temperature well below 10$^{10}$\,K, and a
low-frequency spectral cutoff steeper than $\alpha=3$ if expressed as a
power law. As first reported in Kellermann et al.\ (1999), we find that
the only plausible explanation is free-free absorption from ionised
gas; a conclusion also reached with other data by Kameno et al.\ (2001)
and Kadler et al.\ (2001a). At about 0.5\,pc along the eastern jet, and
1 to 2\,pc along the western jet, the signature of free-free absorption
becomes hard to distinguish from synchrotron-self-absorbed spectral
shapes. If conditions are such that similar peak frequencies result,
the two absorption mechanisms are difficult to disentangle, but it is
likely that both play a role.

The absorption is more pronounced along the western jet than at
corresponding locations along the eastern jet. This suggests a
geometrically thick disk- or torus-like absorbing region, more or less
perpendicular to the jets, which are oriented close to the plane of the
sky, with the eastern jet approaching and the western jet
receding. Variations on this scenario with a thinner disk or torus,
either not oriented orthogonal to the jets, or warped, are
possible. Furthermore, the range of spectral shapes and the flux
density variations of components tracked over time at 15\,GHz
(Fig.~\ref{ffflux}, right panel), suggest that the absorbing region has
a fairly well-defined overall geometry but substantial patchiness in
detail. It is then not very meaningful to use a single epoch for
detailed modeling. The deepest absorption seen, $\tau\sim1$ at 43\,GHz
over the central region, implies a volume density of $n_{\rm
e}\sim10^5$\,cm$^{-3}$ if the free-free absorbing gas were distributed
uniformly along a path-length of 0.5\,pc with a temperature
$T=10^4$\,K.

\begin{figure}[ht!]
\begin{center}
    \null\vskip 0.001 truecm\null\hskip -7.9 truecm
    \psfig{file=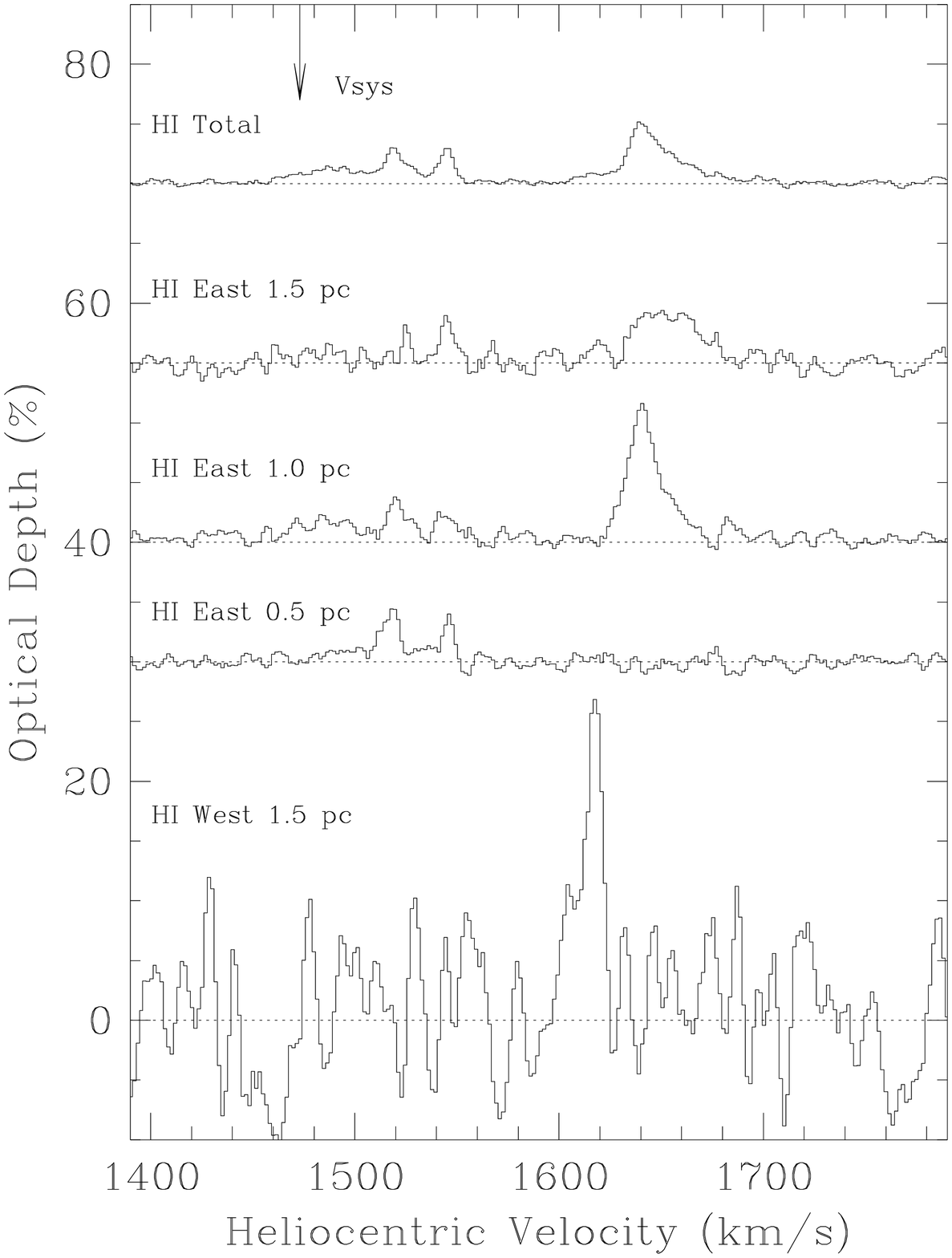,height=9.5cm,angle=0}
    \null\vskip -10 truecm\null\hskip 7.8 truecm
    \psfig{file=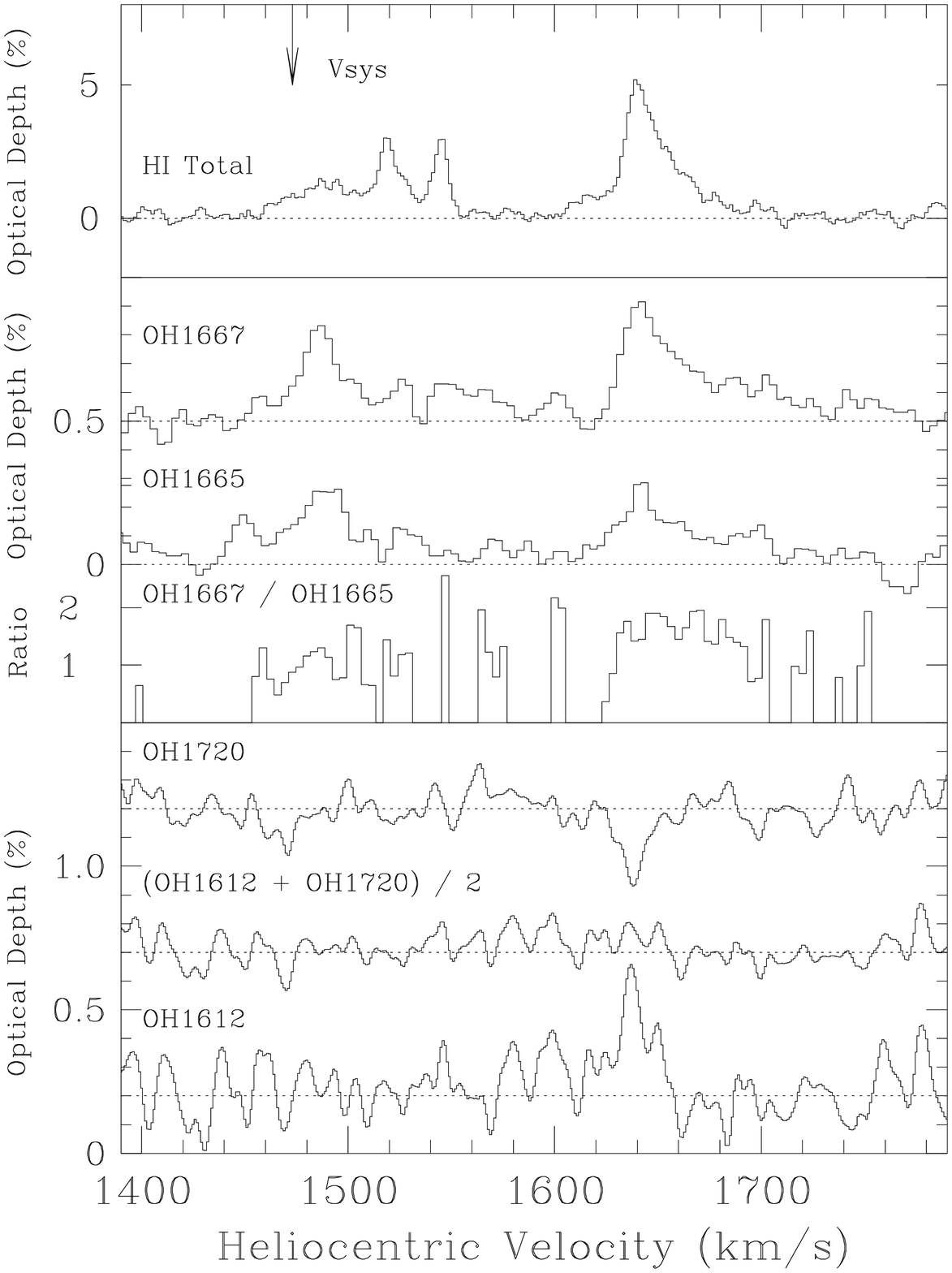,height=9.5cm,angle=0}
    \caption{
        Left: H{\sc i} optical
           depths observed in 1998 at various locations along the jets of
           NGC\,1052. Multiple offsets
           are used for clarity; dotted lines show the zero levels.\hfill\break
        Right:WSRT optical depth measurements (negative is emission)
          for all four 18\,cm OH lines. The offset zero lines are indicated; also
          note the different scale for our integrated VLBI H{\sc i} spectrum,
          shown convolved to the same resolution for comparison.
            }
\label{spectra}
\end{center}
\end{figure}

\section{Atomic gas}\label{atomi}

Figure~\ref{spectra} depicts H{\sc i} VLBI spectra from July 1998 at
all locations where the signal-to-noise ratio allowed detection of
absorption at the level of a few percent. 
We think there are three
different absorption systems, at least two of which are probably due to
atomic gas on parsec or sub-parsec scales, local to the AGN
environment.

The most remarkable system is at ``high velocity'', redshifted by 125
to 200\,km\,s$^{-1}$ with respect to the stellar systemic velocity. 
It gives rise to 
5 to 20\,\%\ 
absorption
at 1 to 2\,pc along both
the western, receding jet, and the eastern, approaching jet, with a
possible west-to-east velocity gradient of some
10\,km\,s$^{-1}$\,pc$^{-1}$. But the high velocity system is absent
closer to the core, at least on the eastern side, where it could easily
have been detected.
The combination of free-free absorption covering the inner parsec, and
atomic gas in an annulus at 1 to 2\,pc, is quite natural: the innermost
region and/or the surface of an accretion disk or torus receive the
most intense ionising radiation. Weaver et al.\ (1999), Guainazzi \&\
Antonelli (1999), Guainazzi et al.\ (2000), and Kadler et al.\ (2002b)
discuss soft X-ray absorbing gas towards the nuclear continuum source,
with large column depths ($N_{\rm H}=10^{23}$ to 10$^{24}$\,cm$^{-2}$),
and possibly a patchy distribution, or involving two components with a
substantially different density, which matches the evidence we see for
patchiness in the radio spectra and 
flux density evolution of jet
components. An H{\sc i} optical depth of 20\%\ with a FWHM of
20\,km\,s$^{-1}$ implies a column depth of $N_{\rm H}=10^{21} T_{\rm
sp,100}$\,cm$^{-2}$, but 
close to an AGN 
the spin temperature $T_{\rm sp}$ 
may well 
be
one or
two orders of magnitude 
above 
100\,K
(e.g., Maloney et al.\ 1996).

\section {Molecular gas}\label{molec}

Figure~\ref{spectra} shows WSRT spectra of the 18\,cm OH lines. We find
that 1667 and 1665\,MHz absorption, recently detected by Omar et al.\
(2002), extends over at least as wide a velocity range as H{\sc i}. The
peak 1667\,MHz depth, 0.4\,\%\ in the high velocity system, suggests a
column depth of order $10^{14}$\,cm$^{-2}$. The 1667/1665 ratio ranges
from near 1 at low velocities to approximately 2 in the high velocity
system. We have discoverd that the satellite lines are also present:
1612\,MHz in absorption and 1720\,MHz in emission. Their conjugate
profiles probably result from excitation in a far infra-red radiation
field when the OH column density is sufficiently large; competing
pumping mechanisms determine which line is in emission and which one in
absorption in specific density and temperature regimes, as modeled for
Cen\,A by van Langevelde et al.\ (1995).

The OH main lines and the total H{\sc i} profile are remarkably similar
in the high velocity system, and we suggest co-location of these high
velocity atomic and molecular gas components. They probably do not
coincide with the H$_2$O masers at 0.1 to 0.2\,pc along the receding jet
(Claussen et al.\ 1998), even though these are at the same
velocity. Other questions also remain regarding the nature of the high
velocity system. Interpretation of the velocity gradient in H{\sc i} as
evidence for a structure rotating around the nucleus is contradicted by
the fact that the centroid is redshifted by 150\,km\,s$^{-1}$ or more
from the systemic velocity. But if it is instead infalling gas, the
nature of the central hole in H{\sc i} is unclear.

\section*{References}








\reference Claussen, M.J., Diamond, P.J.,
   Braatz, J.A., Wilson, A.S., \and Henkel, C. 1998, \apj, 500, L129
\reference de Vries, W. H., Barthel, P. D., \and O'Dea,
C. P. 1997, A\&A, 321, 105
\reference Guainazzi, M., \and
   Antonelli, L. A. 1999, \mnras, 304 L15
\reference Guainazzi, M., Oosterbroek, T.,
   Antonelli, L. A., \and Matt, G. 2000, A\&A, 364, L80
\reference Gabel, J. R., Bruhweiler, D. M.,
   Crenshaw, D. M., Kraemer, S. B., \and Miskey, C. L. 2000, \apj, 532, 883
\reference Heeschen, D. S., \and Puschell, J. J. 1983, \apj, 267, L11
\reference Kadler, M., Ros, E., Kerp, J., et al.
2002a, 
in Proc.\ 6th Eur.\ VLBI Network Symp.,
ed.\ E. Ros
et al.\ 
(MPIfR, Bonn, Germany), 167
\reference Kadler, M., Ros, E., Zensus, J.A., Lobanov, A.P., \and
Falcke, H. 2002b, 
in SRT: the
impact of large antennas on Radio Astronomy and Space Science, SRT Conference
Proceedings, (Cagliari, Italy) Vol.\ 1, in press
\reference Kameno, S., Sawada-Satoh, S. Inoue, M., Shen,
   Z.-Q., \and Wajima, K. 2001, PASJ, 53, 169
\reference Kellermann, K.I., Vermeulen, R.C., Cohen, M.H., \and Zensus,
  J.A. 1999, \baas, 31, 856
\reference Maloney, P. R., Hollenbach, D. J., \and Tielens
   A. G. G. M. 1996, \apj, 466, 561
\reference Omar, A., Anantharamiah, K. R.,
   Rupen, M., \and Rigby, J. 2002, A\&A, 381, L29
\reference Sargent, W.L.W., Schechter, P.L., Boksenberg, A., \and Shortridge, K. 1977, \apj, 212, 326
\reference van Langevelde H.J., van Dishoeck E.F., Sevenster M., \and
   Israel, F.P. 1995, \apj, 448 L123
\reference Vermeulen, R. C., Ros, E., Kellermann, K. I.,
  Cohen, M. H., Zensus, J. A., \and van Langevelde, H.J. 2002, A\&A, submitted
\reference Weaver, K. A., Wilson, A. S.,
   Henkel, C., \and Braatz, J. A. 1999, \apj, 520, 130
\reference Wrobel, J. 1984, \apj, 284, 531


\end{document}